\documentclass[aps,prl,reprint,groupedaddress,superscriptaddress,longbibliography]{revtex4-1}

\usepackage{amsmath}
\usepackage{amssymb}
\usepackage{graphicx}
\usepackage{bm}
\usepackage{epstopdf}
\usepackage{color}
\usepackage{ulem}
\usepackage{endnotes}
\usepackage{eucal}
\usepackage[dvipsnames]{xcolor}
\usepackage{multirow}

\usepackage{float}
\usepackage{array}
\usepackage{booktabs}

\usepackage{graphicx}
\usepackage[caption=false]{subfig}
\usepackage{color}
\usepackage{amsfonts}
\usepackage{lipsum}
\usepackage[super]{nth}
\usepackage[dvipsnames]{xcolor}

\newcommand{\ignore}[1]{}

\begin{document}

\title{Experimental sensing quantum atmosphere of a single spin}

\author{Kehang Zhu}
\thanks{These two authors contributed equally}
\affiliation{Hefei National Laboratory for Physical Sciences at the Microscale and Department of Modern Physics, University of Science and Technology of China, Hefei 230026, China}
\affiliation{CAS Key Laboratory of Microscale Magnetic Resonance, University of Science and Technology of China, Hefei 230026, China}
\affiliation{Synergetic Innovation Center of Quantum Information and Quantum Physics, University of Science and Technology of China, Hefei 230026, China}

\author{Zhiping Yang}
\thanks{These two authors contributed equally}
\affiliation{Hefei National Laboratory for Physical Sciences at the Microscale and Department of Modern Physics, University of Science and Technology of China, Hefei 230026, China}
\affiliation{CAS Key Laboratory of Microscale Magnetic Resonance, University of Science and Technology of China, Hefei 230026, China}
\affiliation{Synergetic Innovation Center of Quantum Information and Quantum Physics, University of Science and Technology of China, Hefei 230026, China}

\author{Qing-Dong Jiang}
\email{qingdong.jiang@fysik.su.se}
\affiliation{Department of Physics, Stockholm University, Stockholm SE-106 91 Sweden}

\author{Zihua Chai}
\affiliation{Hefei National Laboratory for Physical Sciences at the Microscale and Department of Modern Physics, University of Science and Technology of China, Hefei 230026, China}
\affiliation{CAS Key Laboratory of Microscale Magnetic Resonance, University of Science and Technology of China, Hefei 230026, China}
\affiliation{Synergetic Innovation Center of Quantum Information and Quantum Physics, University of Science and Technology of China, Hefei 230026, China}

\author{Zhijie Li}
\affiliation{Hefei National Laboratory for Physical Sciences at the Microscale and Department of Modern Physics, University of Science and Technology of China, Hefei 230026, China}
\affiliation{CAS Key Laboratory of Microscale Magnetic Resonance, University of Science and Technology of China, Hefei 230026, China}
\affiliation{Synergetic Innovation Center of Quantum Information and Quantum Physics, University of Science and Technology of China, Hefei 230026, China}

\author{Zhiyuan Zhao}
\affiliation{Hefei National Laboratory for Physical Sciences at the Microscale and Department of Modern Physics, University of Science and Technology of China, Hefei 230026, China}
\affiliation{CAS Key Laboratory of Microscale Magnetic Resonance, University of Science and Technology of China, Hefei 230026, China}
\affiliation{Synergetic Innovation Center of Quantum Information and Quantum Physics, University of Science and Technology of China, Hefei 230026, China}

\author{Ya Wang}
\affiliation{Hefei National Laboratory for Physical Sciences at the Microscale and Department of Modern Physics, University of Science and Technology of China, Hefei 230026, China}
\affiliation{CAS Key Laboratory of Microscale Magnetic Resonance, University of Science and Technology of China, Hefei 230026, China}
\affiliation{Synergetic Innovation Center of Quantum Information and Quantum Physics, University of Science and Technology of China, Hefei 230026, China}

\author{Fazhan Shi}
\affiliation{Hefei National Laboratory for Physical Sciences at the Microscale and Department of Modern Physics, University of Science and Technology of China, Hefei 230026, China}
\affiliation{CAS Key Laboratory of Microscale Magnetic Resonance, University of Science and Technology of China, Hefei 230026, China}
\affiliation{Synergetic Innovation Center of Quantum Information and Quantum Physics, University of Science and Technology of China, Hefei 230026, China}

\author{Changkui Duan}
\affiliation{Hefei National Laboratory for Physical Sciences at the Microscale and Department of Modern Physics, University of Science and Technology of China, Hefei 230026, China}
\affiliation{CAS Key Laboratory of Microscale Magnetic Resonance, University of Science and Technology of China, Hefei 230026, China}
\affiliation{Synergetic Innovation Center of Quantum Information and Quantum Physics, University of Science and Technology of China, Hefei 230026, China}

\author{Xing Rong}
\email{xrong@ustc.edu.cn }
\affiliation{Hefei National Laboratory for Physical Sciences at the Microscale and Department of Modern Physics, University of Science and Technology of China, Hefei 230026, China}
\affiliation{CAS Key Laboratory of Microscale Magnetic Resonance, University of Science and Technology of China, Hefei 230026, China}
\affiliation{Synergetic Innovation Center of Quantum Information and Quantum Physics, University of Science and Technology of China, Hefei 230026, China}

\author{Jiangfeng Du}
\email{djf@ustc.edu.cn }
\affiliation{Hefei National Laboratory for Physical Sciences at the Microscale and Department of Modern Physics, University of Science and Technology of China, Hefei 230026, China}
\affiliation{CAS Key Laboratory of Microscale Magnetic Resonance, University of Science and Technology of China, Hefei 230026, China}
\affiliation{Synergetic Innovation Center of Quantum Information and Quantum Physics, University of Science and Technology of China, Hefei 230026, China}

\begin{abstract}
Understanding symmetry-breaking states of materials is a major challenge in the modern physical sciences. Quantum atmosphere proposed recently sheds light on the hidden world of these symmetry broken patterns. But the requirements for exquisite sensitivity to the small shift and tremendous spatial resolution to local information pose huge obstacles to its experimental manifestation. In our experiment, we prepare time-reversal-symmetry conserved and broken quantum atmosphere of a single nuclear spin and successfully observe their symmetry properties. Our work proves in principle that finding symmetry patterns from quantum atmosphere is conceptually viable. It also opens up entirely new possibilities in the potential application of quantum sensing in material diagnosis.
\end{abstract}

\maketitle

Symmetry-breaking state of materials is the wellspring of novel and deep physical thoughts \cite{haldane2017}. These subtle forms of symmetry breaking often connect with topology and entanglement \cite{wen_colloquium:_2017,gingras_quantum_2014}. They display a rich variety of phenomena from quantum phase transition \cite{Sachdev} to topological states of matter \cite{hasan}. 
However, to experimentally identify symmetries of a quantum state is extremely non-trivial. For example, transport measurements can end up with trivial results due to formation of domains in a sample \cite{martin_observation_2008}.

\begin{figure}
\centering
\includegraphics[width=1\columnwidth]{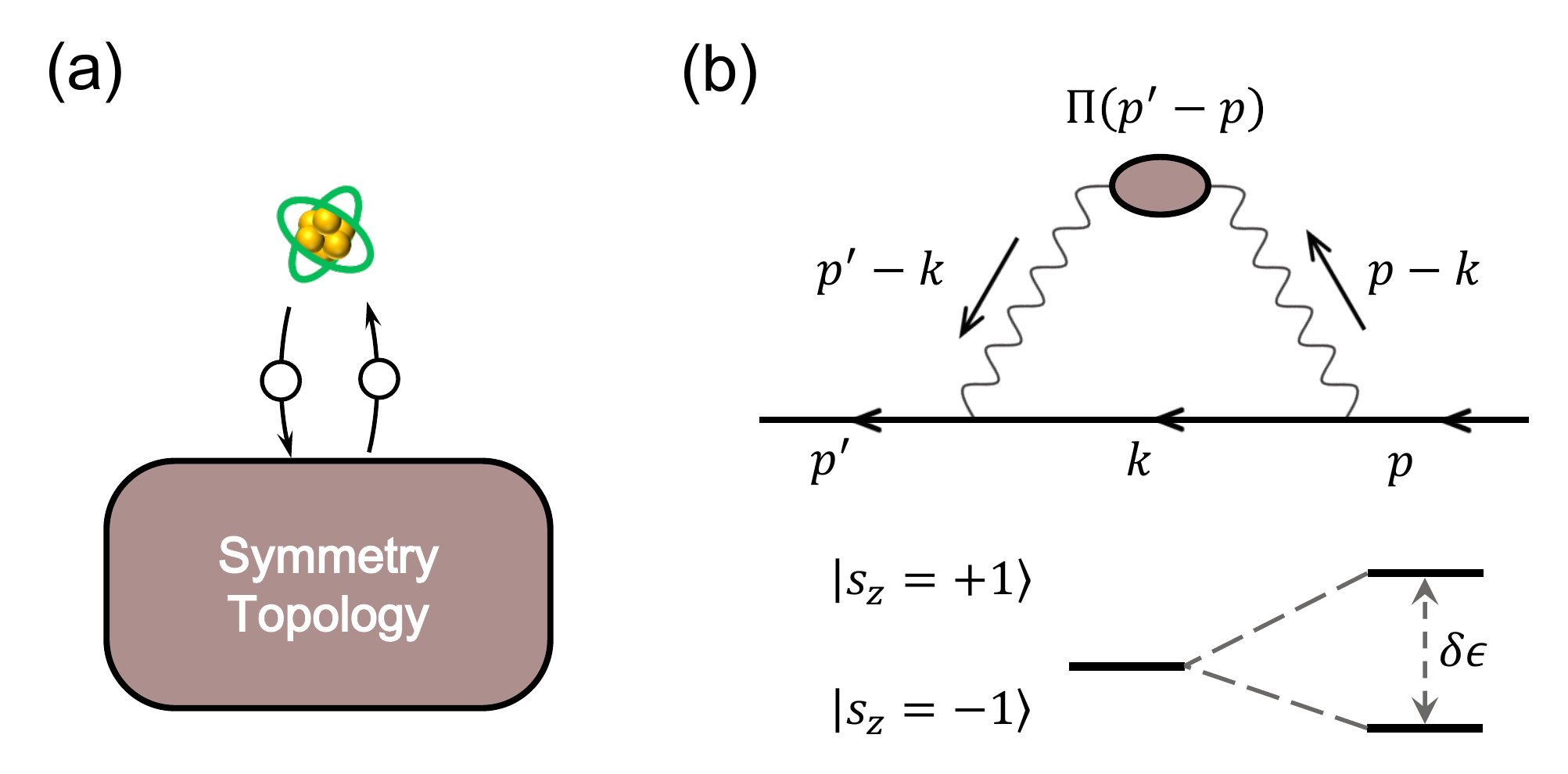}
\caption{\label{figure1} \textbf{Concepts  of quantum atmosphere.} (a) Illustration of quantum atmosphere near the material's surface. The brown box represents domain with hidden symmetry and topology in the material. Due to quantum fluctuation (two-particle-exchange process), local symmetry can be mapped out via the QA near the surface. The circles represent virtue particles created by quantum fluctuation. (b) The physical mechanism of the quantum atmosphere. The upper panel shows a generic Feynman diagram for the QA probe. The spectra of the sensor is modified due to the two-photon-exchange process. The black straight (wavy) line with an arrow represents propagator for sensor spin (photons). The grey circle represents the scattering vertex $\Pi(p\prime-p)$ where the symmetry information in a material is encoded. The lower panel shows that the energy level of the spin-up state and spin-down state will split with an energy difference $\delta \epsilon$ when a spin is put in a time-reversal-symmetry broken atmosphere.}
\label{fig:qa}
\end{figure}

Traditional spectroscopic experiments play prominent roles in characterizing solid structure or spectra of quasi-particles in materials. Over the past century, a lot of techniques have been developed to unravel the symmetry pattern of the materials, i.e., photon scattering \cite{Bennett_x_neutron_2010}\cite{Kim_x_ray_2016}\cite{Stern_x_ray_2014}, neutron scattering spectroscopy \cite{Bennett_x_neutron_2010}\cite{Adams_neutron_2018}\cite{Li_neutron_2019}, electron diffraction\cite{Tom_1927}\cite{Ma_2014} and transverse field muon-spin rotation spectroscopy \cite{Aoki_muon_2003}\cite{Shang_muon_2019}\cite{Zhang_muon_2019}. However, the symmetry of a quantum state can hardly be revealed directly from traditional spectroscopic measurements based on real particle interaction [see supplementary
text SM1 and  Table.S1 \cite{SI}].
More recently, the proposal of quantum atmosphere (QA) has enabled a novel way to directly probe symmetry properties of the materials \cite{jiang_quantum_2019}. QA sensing is based on the idea that quantum fluctuations can bring the symmetry and topology information of a material to its vicinity zone (atmosphere) (Fig. \ref{fig:qa}A). Subsequently, the spectra of a sensor in the atmosphere zone will be disturbed, leaving a fingerprint of certain broken symmetry of that material. The physical mechanism of the QA sensing is based on a virtual-particle exchange process between the sensor and the nearby part of a material (Fig. \ref{fig:qa}B and SM2). Therefore, sensing QA is proposed to be an effective way to map out the topography of symmetry in materials, which is not available in either transport measurements or traditional spectroscopic measurements. Yet, no experiment has been performed to realize its potential. Because QA sensing requires exquisite sensitivity to capture the extremely small electromagnetic signal and tremendous spatial resolution to detect the local information.

Here, we report a method to detect the QA of a single spin and directly observe its symmetry property. The target spin is prepared to different types of atmosphere, which leaves a unique fingerprint of its symmetry in the nearby sensor spin. Since the effects of quantum fluctuations are more dramatic in low dimensions \cite{haldane2017},
our unambiguous demonstration of QA sensing turns this purely theoretical proposal into realistic physical observables, opening up entirely new possibilities in fundamental studies in symmetry detection. The method can be further extended to investigate time reversal symmetry pattern in magnetic materials and parity symmetry in electric dipole systems, paving the way for potential applications of quantum sensing in material diagnosis.

Symmetry information can be directly revealed by measuring the proper physical quantities in the atmosphere.
In the vicinity of the target spin, one can visualize the magnetic field $\delta B$ or magnetic field fluctuation $\delta B^2 $ of the spin. These can be understood from atmospheric point of view (SM3). In the presence of the atmosphere, the free energy for the sensor spin nearby is
\begin{equation}
\mathcal{F}=A_0(\vec I-P I_{z})^2+A_{ij}I_i s_j .
\end{equation}

Here, $\vec s$ stands for the spin of the quantum sensor and $\vec I$ represents the target spin. The polarization direction is chosen to be $\vec z$. $P$ is the polarization of the spin. The coupling constant between the two spins is $A_{ij}$ ($i,j \in \{x,y,z\}$). The stiffness of the target spin's magnetic field, $A_0$, is inverse to the spin orientation fluctuation rate. In the case of totally polarized target spin, $A_0 \rightarrow \infty$, so fluctuation is forbidden and $\vec I$ is restricted to $P I_{z}$. In the case of unpolarized target spin, $A_0$ is finite, which indicates that fluctuation is very large, and there is no favorable direction of the target spin.
The fluctuation part of the target spin can be integrated out to yield the effective free energy of the quantum sensor in the atmosphere of the target spin (SM3).
Considering the longitudinal coupling $A_{zz}$ far exceeds other coupling components, we get
\begin{equation}
\mathcal F_{eff}=A_{zz}PI_{z}s_z+\frac {A_{zz}^2}{4A_0}s_z s_z .
\end{equation}
When the target spin is polarized $(P\neq0)$, the quantum sensor is put in a time-reversal symmetry broken atmosphere and the first term is allowed. In this case, the mean magnetic field $\delta B \neq 0$. Thereby, the energy level of the spin-state will split with an energy difference $\delta \epsilon$ as shown in Fig. \ref{fig:qa}b. When the target spin is unpolarized ($P = 0$), time-reversal symmetry is conserved and only the second term can exist. Based on this effective action, $\left< s \right >=0$ while $\left < s_zs_z \right > \neq 0$, which indicate that $ \delta B= 0$ and the magnetic field fluctuation $\delta B^2 \neq 0$.

\begin{figure}[http]\centering
\includegraphics[width=1.1\columnwidth]{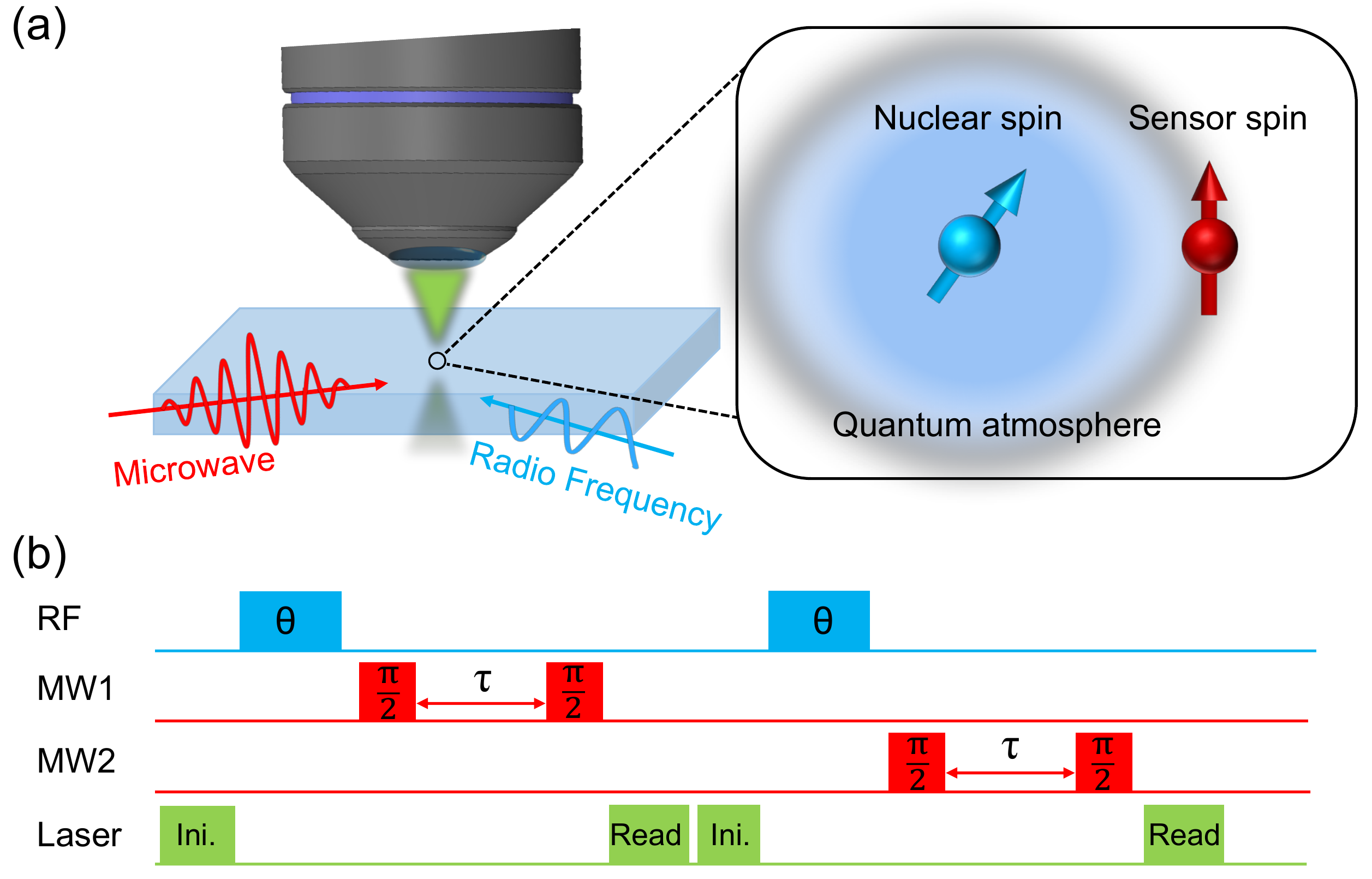}
\caption{\label{figure2} \textbf{Schematic of the setup and experimental method.} (a) Nitrogen-vacancy centre in the quantum atmosphere of a nearby $^{13}\mathrm{C}$ nuclear spin is illuminated by a focused green laser beam and controlled by microwave and RF pulses. In the magnified figure, red ball with an arrow represents the target spin and the blue one is the sensor spin. 
(b) Method of preparation: The sensor spin and target spin are first initialized by laser pumping.
Radio-frequency pulses are used to transfer the population between nuclear spin's states.
And by varying the rotation angle of RF ($\theta$), we can prepare
distinct $^{13}\mathrm{C}$ nuclear spin's quantum atmospheres from time-reversal symmetry conserved type to broken ones.
Detection (Ramsey sequence): the sensor spin is prepared in the $(\left | 0\right >+i\left | -1 \right >)/\sqrt 2$ state with a microwave (MW)  $\pi/2$  pulse along the $\hat x$ axis. Subsequently, within the sensing process, the spin evolves under the influence of nuclear spin's quantum atmosphere for duration $\tau$, immediately followed by another microwave  $\pi/2$ pulse along the same axis. Then the spin polarization is read out by the laser.
Microwave pulses, MW1 and MW2, are used for recording the sensor spin's evolution in the different $^{13}\mathrm{C}$ states.}
\label{fig:setup}
\end{figure}

We experimentally demonstrate our proposal with a nitrogen-vacancy (NV) center in diamond \cite{jelezko_observation_2004}\cite{childress_coherent_2006} with a nearby $^{13}\mathrm{C}$ nuclear spin. An NV-based optically detected magnetic resonance (ODMR) setup ( (Fig. \ref{fig:setup}A and SM4) is constructed to detect $^{13}\mathrm{C}$'s quantum atmosphere. Under the external magnetic field of 515 Gauss, the energy levels of NV's electron spin $\left | S_z = 0 \right >$ and $\left | S_z = -1 \right >$ are selected to realize an effective two-level quantum sensor (sensor spin), which can be initialized by laser pulses and manipulated by microwave (MW) pulses to capture the magnetic quantities in the target spin's atmosphere. The state of the target spin can be engineered by radio-frequency (RF) pulses.

The Hamiltonian of the effective two-level system is
\begin{equation}
H =  DS_z+ \gamma _e \vec B_{ext}\cdot \vec{ S}+ \gamma _e  \delta \vec B\cdot \vec {S},  
\end{equation}
where S is Pauli spin-1/2 operator of the two-level subspace spanned by the spin states $\left | S_z = 0 \right >$ and $\left | S_z = -1 \right >$, $B_{ext}$ is the external magnetic field applied along NV's axis, D is the zero filed splitting between sensor spin's $\left | S_z = 0 \right >$ and $\left | S_z = -1 \right >$ states, which is measured as $2.87$ GHz, $\gamma_e $ is electron's gyromagnetic ratio being $2.803~\mathrm {MHz/Gauss}$, $\delta B $ is the magnetic field the sensor detects.
In the rotation frame, the sensor's Hamiltonian is
$H_{rot} =A_{zz}I_z {S_z} $ (SM5), where $A_{zz}$ is the longitudinal coupling between the sensor and the nuclear spin being $13.56$ MHz in our experiment.
Other components can be neglected because they are much weaker than the longitudinal component.
Therefore, the magnetic field is measured as $\delta B=\left <A_{zz}PI_z\right >/\gamma_e $ and magnetic fluctuation as $ \delta B^2 = \left < (A_{zz}(I_z - PI_z))^2\right >/\gamma_e^2 $. The bracket means the average of repetitive measurements.

To probe the quantum atmosphere of the nuclear spin, we monitor the sensor spin's evolution under the influence of $^{13}\mathrm{C}$ which has different polarization along the $+\hat z$ direction.
The sensor spin is first initialized by laser pumping. In the meanwhile, excited-state level-anticrossing near 510 Gauss allows electron-nuclear-spin flip-flops to occur near resonantly \cite{Steiner_excited_2010}. Since $^{13}\mathrm{C}$ preferentially relaxes into $\left | m_I=\uparrow \right >$ during the process, high degree of initial polarization is prepared.
RF pulses are then applied to transfer the population between the $\left | m_I=\uparrow \right >$ and $\left | m_I=\downarrow \right >$ states (Fig. \ref{fig:setup}B and SM6) \cite{xu_dynamically_2018}.
Different $^{13}\mathrm{C}$ nuclear spin states corresponding to distinct quantum atmospheres can be prepared by controlling the rotation angle ($\theta$) of RF pulses.
 It follows with a Ramsey sequence \cite{balasubramanian_ultralong_2009} to record magnetic field information from $^{13}\mathrm{C}$ nuclear spin.
The sensor spin is initially rotated to $(\left | 0\right >+i\left | -1 \right >)/\sqrt 2$ state with a microwave $\pi/2$  pulse along the $\hat x$ axis.
During the sensing process, the sensor spin evolves under the influence of nuclear spin's quantum atmosphere for a duration $\tau$. Subsequently, another microwave $\pi/2$ pulse along the same axis is applied on the sensor spin followed by the final read-out of the spin polarization by the laser.
Limited by the power of the microwave (pulse excitation bandwidth 6.5 MHz), the whole magnetic field information is unavailable in a single Ramsey sequence.
Therefore, we adopt the strategy to split the sensing process into two steps to record the sensor spin's evolution in the different $^{13}\mathrm{C}$ states.
In the first step, MW1 is detuned from the sensor spin's resonance frequency between $\left | m_s=0, m_I=\uparrow \right >$ and $\left | m_s=1, m_I=\uparrow \right >$ by +1MHz (4321.0 MHz).
The large off-resonance of the sensor spin in the $\left | m_I=\downarrow \right >$ subspace leaves only the information of $\left | m_I=\uparrow \right >$ state observable.
Complementarily, the procedure above is repeated but with MW2 detuned from the resonance frequency between $\left | m_s=0, m_I=\downarrow \right >$ and $\left | m_s=1, m_I=\downarrow \right >$ by $-$1MHz (4305.5 MHz ).
The experiment has been repeated for $6\times 10^5$ times to build good statistics.

\begin{figure*}[http]\centering
\includegraphics[width=2\columnwidth]{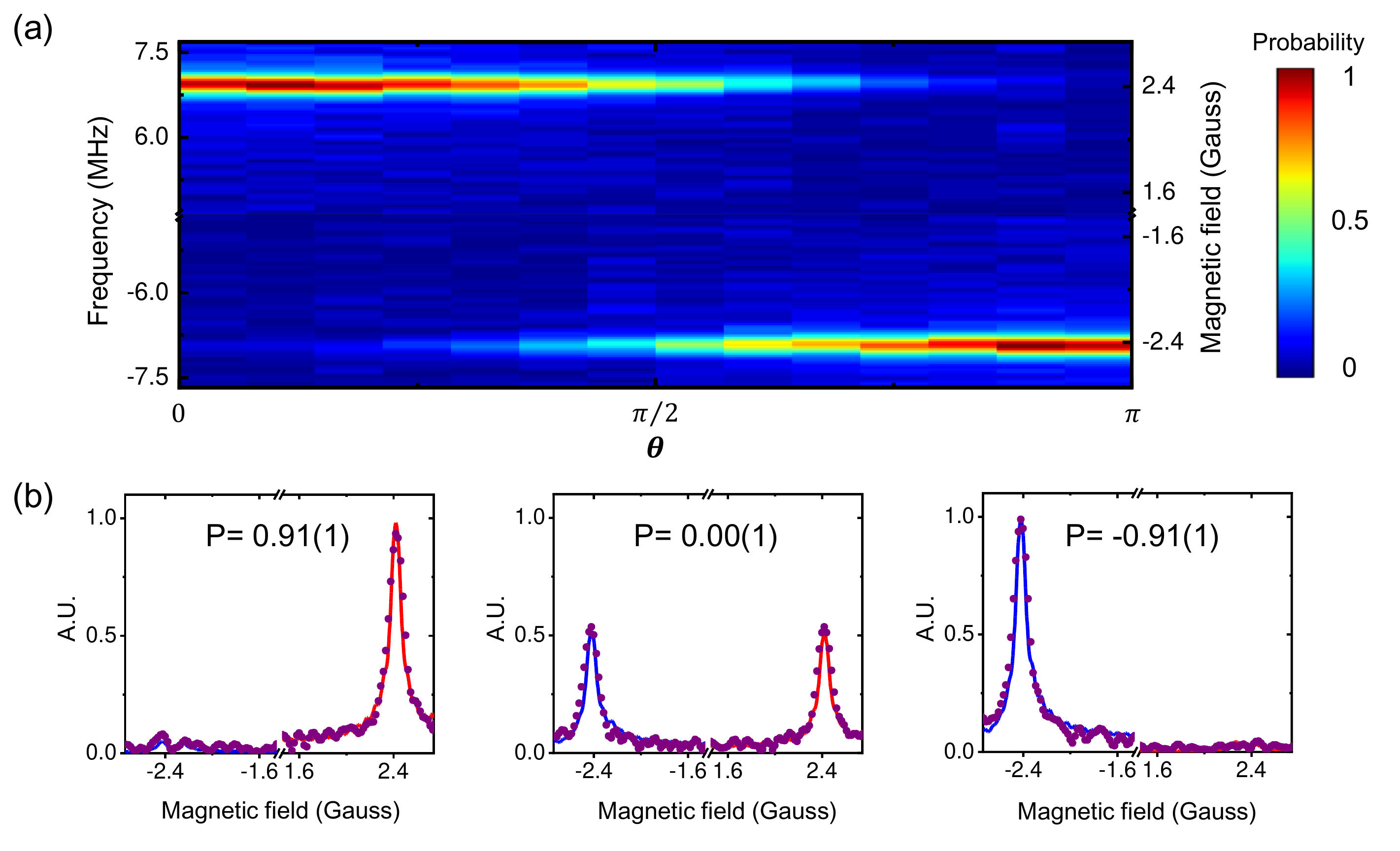}
\caption{\label{figure3} \textbf{Spectrum of quantum atmosphere with varying symmetry property.} (a) The whole spectrum of fifteen states polarized to different quantum atmospheres with different RF rotation angle. The brighter the color is, the more likely the magnetic field will distribute at that field strength. (b) three typical quantum atmospheres.  The nuclear spin polarization is $P=P_{0} cos(\theta)$. Left: The nuclear spin is polarized to its $ \left | m_I=\uparrow \right >$ state corresponding to a time-reversal symmetry broken atmosphere, it applies a positive magnetic field to the sensor spin. Middle: The nuclear spin is fully unpolarized within the margin of error, no magnetic field is applied to the sensor corresponding to a time-reversal symmetry conversed atmosphere. Right: the nuclear spin is polarized to its $\left | m_I=\downarrow\right > $ state, it applies a negative magnetic field to the sensor corresponding to a time-reversal symmetry broken atmosphere. The scattered dots are the experimental results while the real lines are our simulant results. Every point is averaged over $6 \times 10^5$ repetitive measurements.
 }\label{fig:spec}
\end{figure*}

During the sensing process, the sensor spin state undergoes Larmor precession at different frequencies due to the influence of $^{13}\mathrm{C}$ nuclear spin, which is in a superposition state of $\left | m_I=\downarrow \right >$ and $\left | m_I=\uparrow \right >$. 
The two eigenstates exert different magnetic fields on the sensor, shifting its original Larmor frequency $f_0$ to $f_0- A_{zz}/2$ and $f_0+A_{zz}/2$.
By varying the duration $\tau$, the evolution of the sensor spin can be recorded in the time domain.
To obtain the effective magnetic field, we transform the signal from time domain to frequency domain and divide it by electron's gyromagnetic ratio.

 Fig. \ref{fig:spec}A depicts the observed phase transition of $^{13}\mathrm{C}$'s quantum atmosphere between time-reversal symmetry conserved type and broken ones at different levels.
 These states are prepared to different polarization, $P$, ranging from $-P_{0}$ to $P_{0}$. The nuclear spin polarization is
$P=P_{0} cos(\theta)$, where $\theta$ is the rotation angle of RF pulse and $P_{0}$, measured to be $0.91(1)$, is the natural polarization due to optical pumping.
The experimentally prepared polarization of target nuclear spin may deviates from the expected value due to several factors, e.g., laser-induced depolarization \cite{jacques_dynamic_2009} and temperature fluctuation.
Fig. \ref{fig:spec}(B-D) shows three representative QAs in Fig. \ref{fig:spec}A and their magnetic field distributions. For nuclear spin with time-reversal symmetry broken atmosphere, its magnetic field is highly concentrated in the positive (Fig. \ref{fig:spec}B with $P=0.91(1) $) or negative (Fig. \ref{fig:spec}D with $P=-0.91(1)$) part centered near $A_{zz}/(2\gamma_e)=\pm 2.419$ Gauss.
While for the time-reversal symmetry conserved one (Fig.\ref{fig:spec}C with $P= 0.00(1)$), its magnetic field is evenly distributed on both sides.

To quantitatively determine the symmetry of the $^{13} C$ nuclear spin's atmosphere, we first measure the mean magnetic field $\delta B$ and magnetic fluctuation $\delta B^2 $ of each polarization state (SM7). The Symmetry Indicator $\Gamma$ is defined as the ratio of the magnetic fluctuation to the square of magnetic field, $\Gamma= {\delta B^2}/{(\delta B)^2}$, which provides a method to quantitatively measure the symmetry-breaking level of quantum atmosphere. When $ \Gamma$ equals 0, the broken level of time-reversal-symmetry reaches maximum (i.e. the spin is fully polarized) and as the parameter would grow larger, the symmetry breaking level gets lesser. Notably, $\Gamma$ comes to divergence when time-reversal symmetry is conserved.

\begin{figure}
\includegraphics[width=1\columnwidth]{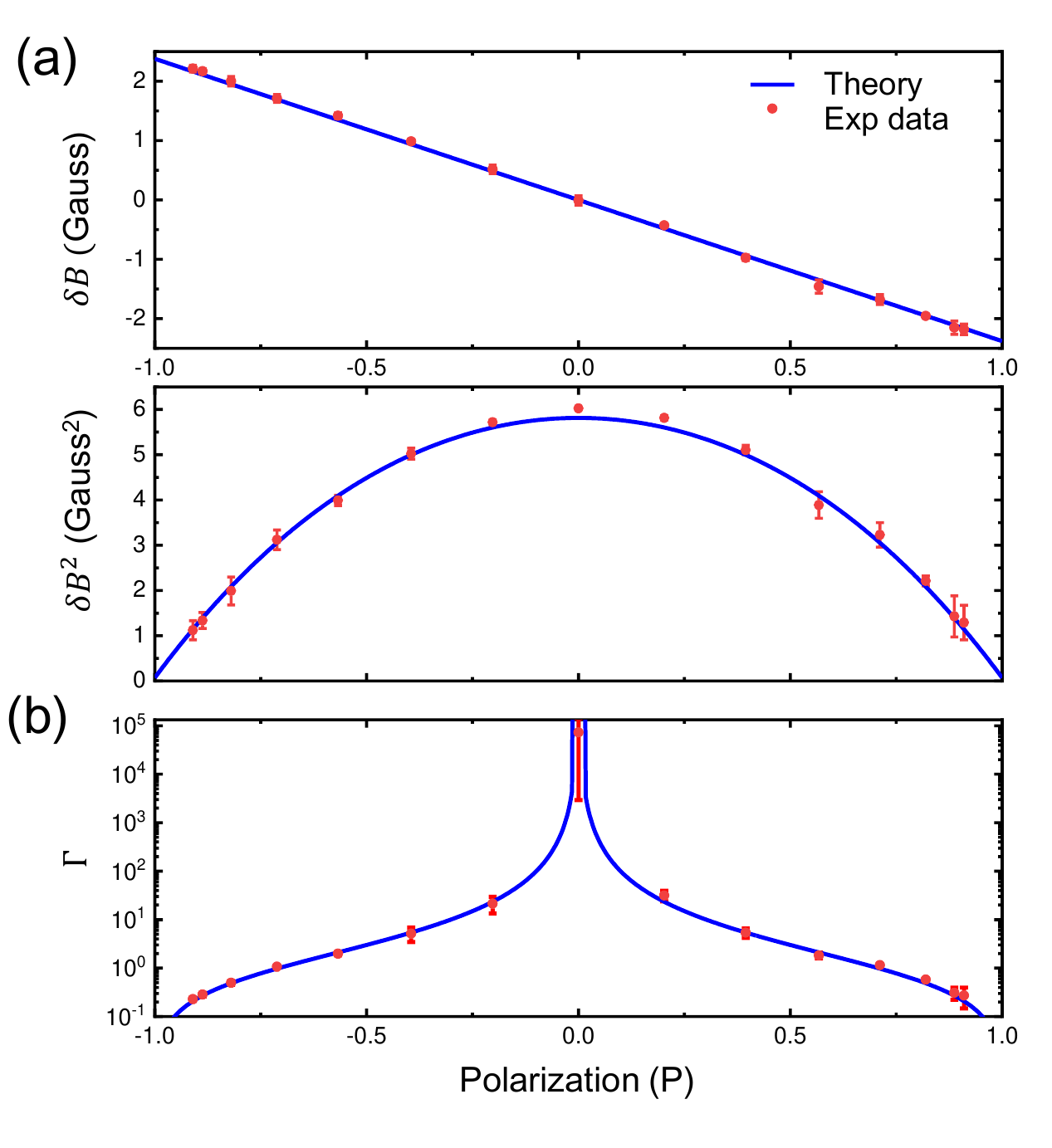}
\caption{\label{figure4} \textbf{ Phase diagram of different quantum atmosphere.} The red dots denote the experiment data, the blue curves denote the theoretical predictions, which we find agreement. (a)The physical observables in the experiment are the mean magnetic field $\delta B$ and the magnetic fluctuation $\delta B^2$ of the $^{13}\mathrm{C}$ spin. The upper panel shows the field of $^{13}\mathrm{C}$'s QA  with different polarization $P$. The lower panel shows their magnetic field fluctuation of $^{13}\mathrm{C}$'s QA . Notably, when time-reversal-symmetry is conserved, there is no favorable direction of the $^{13}\mathrm{C}$ spin, which results in to zero net magnetic field and maximum magnetic field fluctuation. (b) shows symmetry indicator $\Gamma$ as a function of  $P$. $\Gamma$  sees a huge spike around $P=0$ and sharply decreases as $P$  deviates from the that point. It is quite clear from the diagram that time reversal symmetry is broken except the state that nuclear spin is unpolarized. Error bars are mainly due to the photon statistics.
}\label{fig:phase}
\end{figure}

Fig.\ref{fig:phase}A exhibits some realistic physical observables of quantum atmosphere with different time-reversal symmetry broken level. The error bars are obtained by dividing the data into three groups and calculating the variance from their deviations from the mean value. The upper panel shows the magnetic field of QA and the lower panel shows their fluctuation. When the target spin is not polarized, there is no favorable direction. As a result, no magnetic field is exerted by the $^{13}\mathrm{C}$ spin but maximum fluctuation is presented. When it is polarized, the spin shows certain orientation. In this case, even the slightest fingerprint of nonzero energy shift (effective magnetic field) could be sensed due to the exquisite sensitivity of NV center \cite{lovchinsky_nuclear_2016}.
The measured magnetic fluctuation is slightly lifted from the theoretical value (SM7), which is mainly contributed by the inevitable background noise.
Fig.\ref{fig:phase}B shows $\Gamma$ as a function of $^{13}\mathrm{C}$ spin polarization $P$. $\Gamma$  sees a huge spike around $P=0$ and, within the rage of error permitting, $\Gamma$ goes through the divergent point. As $P$  deviates from this point, $\Gamma$ sharply decreases. Judging from the property of the symmetry indicator, we associate this point with the time-reversal-invariant quantum atmosphere and other cases as time-reversal symmetry broken quantum atmosphere \cite{jiang_quantum_2019}.

With the proof-in-principle implementation of QA sensing of a single spin, a clear picture of quantum atmosphere and a method to capture the fingerprint of symmetry are established. A lot of exotic materials could be diagnosed. For example, it is suggested that quantum fluctuation involving Chern-Simons interaction will produce a sort of parity and time-reversal symmetry violating atmosphere above a topological insulator, which induces an effective Zeeman field on the quantum sensor nearby\cite{jiang_quantum_2019}.  Its strength is within NV centers sensitivity of magnetometry. Furthermore, chiral superconductivity may also be directly identified by sensing its time-reversal symmetry violated QA.
Several improvements and further investments are required to extend these real materials diagnosis via QA sensing.
First, ODMR set-up with cryogenic system is necessary to be build. Besides, the influence of the surface impurity needs to be carefully investigated and other magnetic properties of the bulk material near the surface need to be further explored.

Although in this experiment a strongly coupled nuclear spin is employed, it is important to note that the physical picture and general methods are applicable to any target spins irrelevant to its coupling with the sensor\cite{Degen_weakmeas}. For example, $^{13}\mathrm{C}$ spin with a coupling weaker by several orders of magnitude ($A_{zz} \sim$ kHz) has been experimentally accessed and manipulated with the method of high order dynamical decoupling sequence\cite{Jorg_weak}. Adopting specific polarization and readout schemes\cite{Pan_weak}, target spins, weakly and strongly coupled, could be sensed in its quantum atmosphere to reveal their symmetry patterns.

In conclusion, we demonstrate the feasibility of sensing quantum atmosphere using NV-based quantum sensing techniques and turn this purely theoretical concept into realistic physical observables. In the future, our method can be extended to reveal subtle forms of symmetry breaking in materials, opening up entirely new possibilities in the searching and the study of hidden symmetries.
The present observation also raises intriguing possibility in diagnosis more complex materials \cite{jiang_quantum_2019}, e.g., topological insulators and superconductors \cite{hasan}\cite{qi_topological_2011-1}.

\section{Acknowledgments}
We thank Xi Kong for the helpful discussion.
\textbf{Funding:} This work was supported by the National Key R$\&$D Program of China (Grants Nos. 2018YFA0306600, 2016YFB0501603 and 2017YFA0305000), the NNSFC (Nos. 11761131011 and 11775209), the Chinese Academy of Sciences (Grants Nos. GJJSTD20170001, QYZDY-SSW-SLH004 and QYZDB-SSW-SLH005), and Anhui Initiative in Quantum Information Technologies (Grant No. AHY050000). X.R. and F.S. thank the Youth Innovation Promotion Association of Chinese Academy of Sciences for the support. Q.D.J. acknowledges support from the European Research Council (Grant No. 742104) and Swedish Research Council (Contract No. 335-2014-7424).

\renewcommand{\refname}{References and Notes}
\clearpage

\onecolumngrid
\vspace{1.5cm}

\section{Supplementary Material for Experimental sensing quantum atmosphere of a single spin}

%
%
%
%
%
%
%
%
%

\subsection{Comparison between traditional spectroscopy and QA-based spectroscopy}
Traditional spectroscopic experiments play prominent roles in characterizing the crystalline structure or spectra of quasi-particles in materials.
However, most of these spectroscopies are only capable of revealing certain symmetries of the Hamiltonian, but not the symmetry of a specific ground state. Note that the ground state can have less symmetry than its parent Hamiltonian, which is the idea of spontaneous symmetry breaking. Therefore, the symmetry of a quantum state can hardly be revealed directly from traditional spectroscopic measurements.

By sharp contrast, QA sensing bases on the idea of measuring the physical quantities in the material's vicinity zone (atmosphere).  Quantum fluctuations inevitably bring symmetry information of a material into its atmosphere, which could be subsequently measured by a quantum sensor. While we focus on the time-reversal symmetry in this paper, various more quantum sensors could be designed to reveal other kinds of symmetry in principle.

Although the local time-reversal symmetry can be detected with the muon-spin rotation spectroscopy ($\mu$SR). The requirement of a large accelerator to create the $\mu$ particle restricts the application of this technique. Not to mention this type of detection causes damage of the material. Contrarily, QA-based spectroscopy enables revelation of PT-symmetry and other various hidden symmetries. This novel method also offers easy access to the direct diagnosis of diverse materials without damage. The detailed comparisons are shown in the Table. \ref{table1}.

\subsection{general theory for symmetry detection}

The symmetry information of a material can be revealed through a two-virtual-particle exchange process between the sensor and the nearby part of a material.
Consider the general formula for describing interaction between the sensor and nearby material.
The action for the system is
\begin{equation}
S_{mat}=\int d^4x \{\psi_m^\dagger(i\partial_t-H_{mat})\psi_m+\psi_m^\dagger g^\mu A_\mu \psi_m\},
\end{equation}
where the electron-photon coupling term $\psi_m^\dagger g^\mu A_\mu \psi_m$ is included. $H_{mat}$ is the Hamiltonian for the material; $\psi_m$ and $\psi_m^\dagger$ are the Grassmanian fields of electrons; $A_{\mu}$ is the field for photon; $g^\mu$ is a generic operator. Repeated indices means summation in our formulas.

Integrating out Grassmanian fields, one can obtain an effective action for photons,
\begin{equation}
\begin{aligned}
S_{ph}=&-\left\{{\rm{Tr}~ln}\left[(i\partial_t-H_{mat})+g^{\mu}A_{\mu}\right]-{\rm{Tr} ~ln} (i\partial_t-H_{mat})\right\}\\
=&-{\rm Tr~ln}(1+G_0 g^\mu A_\mu)\\
=&-{\rm Tr}(G_0g^\mu A_\mu)-\frac{1}{2}{\rm Tr}(G_0 g^\mu A_\mu G_0g^\nu A_\nu).
\end{aligned}
\end{equation}
In the formula, $G_0=(i\partial_t-H_{mat})^{-1}$ is the Green function for electrons in the material.
By Defining the polarization function $\Pi^{\mu\nu}(p)\equiv \sum_q G_0(q)g^\mu(q)G_0(p-q)g^\nu(p-q)$, the material provides a two-photon vertex
$\Pi^{\mu\nu}A_\mu A_\nu$.

We know that the free electron-photon coupling is $e\psi^\dagger \gamma^\mu A_\mu \psi$. Thus, the additional self-energy induced by the nearby material is
\begin{equation}
\delta \epsilon(p)= e^2\sum_{p^\prime} \Pi(p^\prime)G(p^\prime+p) D(p^\prime)D(-p^\prime),
\end{equation}
where $G(p)$ represents the Green function for the free electron, and $D(p)$ represents the photon propagator in vacuum.
In the derivation above, we set $\hbar=c=1$.

Such a process is generically represented by the Feynman diagram Fig 1b. in the main text.

\subsection{quantum atmosphere of a single spin }

According to the general theory above, the free energy of a single spin is supposed to be written as follows,
\begin{equation}
\mathcal{F}=\frac{1}{2\mu_0}(\vec B-\vec B_0)^2+A \bar \psi (\vec B \cdot \vec \sigma) \psi
\end{equation}
where $(\bar \psi , \psi )$ are the Grassmanian fields for fermions and $\vec \sigma$ are the Pauli operators for spin 1/2 particles. $\vec B$ is the magnetic field, and $\vec B_0=B\hat e_z$ is the applied magnetic field.
In this sense, the spin operator is $\hat s =  \bar \psi \vec \sigma \psi$.
If we define the shift variable $\delta \vec B$ as $\delta \vec B =\vec B-\vec B_0$, the free energy becomes
\begin{equation}
\mathcal{F}=\frac{1}{2\mu_0}(\delta \vec B)^2+A \bar \psi (\vec B_0 \cdot \vec \sigma) \psi+A \bar \psi (\delta \vec B \cdot \vec \sigma) \psi
\end{equation}
Integrating out the shift variable  $\delta \vec B$, one can obtain the effective free energy of the sensor spin.
\begin{eqnarray}
\mathcal F_{eff}&=&A\bar \psi (\vec  B_0 \cdot \vec \sigma) \psi+\mu_0 g^2 \bar \psi(t) \vec {\sigma(t)} \psi(t) \bar {\psi(t)} \vec {\sigma(t)} \psi(t),\\
&=&g \vec B_0\cdot \vec s +\mu_0 g^2 \vec s(t) \vec s(t)
\end{eqnarray}

In order to make the content more understandable, we adopt a more intuitive way in the main text. Instead of complex Grassmanian field, we simply use the spin operator $\vec\sigma$ to denote the fermion (particularity, spin 1/2). In the general theory, we integrate out the fermion to obtain the effective scattering vertex. While in the following, the magnetic fluctuation part is integrated out to yield the effective free energy.

By putting a quantum sensor close to the spin, one can visualize the magnetic field $\delta B$ or magnetic field fluctuation $\delta B^2 $ of the spin. These can be understood from atmospheric point of view. The free energy for the spin of the quantum sensor is
\begin{equation}
\mathcal{F}=\frac{1}{2\mu_0}(\vec B-\vec B_0)^2+g \vec B \cdot \vec s
\end{equation}
Here $\vec s$ is the spin of the quantum sensor and $\vec B$ is the magnetic field, and $\vec B_0=B\hat e_z$ is the applied magnetic field.

In the presence of a target spin $\vec I$, the free energy becomes
\begin{equation}
\mathcal{F}=A_0(\vec I-{P}\vec I_{z})^2+A_{ij}I_i s_j
\end{equation}
where $\vec I$ represents the target spin and $\vec I_0$ represents polarization direction. And the polarization direction is chosen to be $\vec z$. $A_{ij}$ is the coupling constant between the two spins, $A_0$ characterizes the stiffness of the target spin's magnetic field, which is inverse to the spin orientation fluctuation rate. In the case of totally polarized target spin, $A_0 \rightarrow \infty$, so the fluctuation is forbidden and $\vec I$ is restricted to $ {P} I_{z}$. In the case of unpolarized target spin, $A_0$ is finite, which indicates that fluctuation can be very large, and there is no favorable direction of the target spin. Let $\Delta \vec I=\vec I-{P} I_{z}$ represents fluctuation of the target spin, the free energy can be transformed into
\begin{equation}
\mathcal{F}=A_0(\Delta \vec I)^2+A_{ij}(\Delta I_i+ {P}I_{z} s_j)
\end{equation}
The fluctuation part of the target spin can be integrated out, yielding the effective free energy of the quantum sensor in the atmosphere of the target spin:
\begin{equation}
\mathcal F_{eff}=A_{ij} {P}I_{z}s_j+\frac {A_{ij}A_{il}}{4A_0}s_j s_l
\end{equation}
In the system explored below, $A_{zz}$ far exceeds other components.
And considering the fact that $\left <s_xs_z\right >$ and $\left <s_ys_z\right >$ terms are averaged to be zero, we only keep the leading term
\begin{equation}
\mathcal F_{eff}=A_{zz} {P}I_{z}s_z+\frac {A_{zz}A_{zz}}{4A_0}s_z s_z
\end{equation}

\subsection{Sample and setup}

In our experiment we use a [100]-oriented CVD-grown diamond with natural $^{13}$C concentration (1.1\%).  The NV centers were created by $^{15}N^+$ ion implantation at an energy of $5$ keV and subsequent annealing at 1000 $^\circ$C.

Our setup is a home-built optically detected magnetic resonance (ODMR) spectrometer. The pumping 532 nm green laser beam is focused through the oil objective (100 Oil, N.A. 1.25) while the photon sideband fluorescence (wavelength, 650-800 nm) is collected by the same objective.
The pump beam passes through acoustic-optic modulator (AOM) (power leakage ratio 30 dB) twice before reaching the objective. The fluorescence photons are collected by an avalanche photo-diode (dark counts $<$ 100 per second) after passing the optical filters. A permanent magnet provides the external static magnetic field. The magnetic field is aligned by analyzing the splitting between  $\left | S_z = 0 \right >$ and $\left | S_z = -1 \right >$ ($\left | S_z = 0 \right >$ and $\left | S_z = +1 \right >$). A temperature controlled device is used to keep the temperature stable.

The electron spin's evolution is manipulated by the microwave pulses, which are generated by an I/Q modulation with an Arbitrary Waveform Generator (AWG) (CRS1W000B, Chinainstru $\&$ Quantumtech (hefei) Co.,Ltd.).
Two separate channels of the AWG generate the I and Q signal, respectively. These signals combined with another microwave from a RF signal generator (Stanford SG386) are transmitted to the IQ mixer (Marki IQ1545). An Arbitrary Sequence Generator (ASG) (ASG-GT50-C, Chinainstru $\&$ Quantumtech (hefei) Co.,Ltd.) is served to provide TTL signals to control the switch of the AWG, AOM and the time bin of the counter. The radio frequency pulses controlling the nuclear spin are generated by another RF signal generator (Stanford SG386).

\subsection{Spectroscopy of the system}
The Hamiltonian of the whole NV electron spin-$^{13}$C nuclear system is
\begin{eqnarray}
H & = & H_{\textrm{NV}} + H_{\textrm{nuc}} + H_{\textrm{int}}, \\
  & = & DS_z + \gamma_e \textbf{B}\cdot \textbf{S }+ \omega I_{z}+ \textbf{S}\cdot\textbf{A}\cdot\textbf{I},
\end{eqnarray}
where S is Pauli spin-1/2 operator of the two-level subspace spanned by the sensor spin states $\left | S_z = 0 \right >$ and $\left | S_z = -1 \right >$, $B_{ext}$ is the external magnetic field applied along NV's axis, D is the zero filed splitting between sensor spin's $\left | S_z = 0 \right >$ and $\left | S_z = -1 \right >$ states, which is measured as $2.87$ GHz, $\delta B $ is the magnetic field the sensor detects, $\omega$ is the Larmor frequency of the $^{13}C$ spin and $A$ is the hyperfine interaction matrix between the sensor spin and target spin.

In the rotating frame of electron spin, omitting the self-Hamiltonian of NV center, the simplified Hamiltonian reads as,
\begin{eqnarray}
H & = & \Delta S_z+ S_z \left(  A_{zx}I_{x} +  A_{zy}I_{y} +  A_{zz}I_{z} \right)  + \omega_L I_z , \\
  &  \approx & \left(\Delta +  A_{zz}I_{z} \right) S_z + \omega_L I_z,
\end{eqnarray}
where $\Delta$ is the detuning between the driven MW field and the original electron's Larmor frequency.
The strength of the longitudinal coupling $A_{zz}$ between target $^{13}$C nuclear spin and NV center is 13.56 MHz and  the strength of the transverse coupling $\sqrt{A_{xz}^2+A_{yz}^2}$ is 2.8(1) MHz.
Due to $^{13}$C's quantum atmosphere, the frequency shift in the spectroscopy of the NV center is $\pm A_{zz}/2$.
The effective magnetic field interacted NV feels is,
\begin{equation}
B_{eff}  = \frac{A_{zz}}{2\gamma_e} P,
\end{equation}
where P is the polarization of the  $^{13}C$ nuclear spin.

\subsection{Experimental methods}
The  energy level structure of the NV and the $^{13}$C nuclear system is show in (Fig. \ref{figure1}). The triplet ground state has a zero-field splitting D (2.87 GHz) separating the state $| 0 \rangle $ and the degenerate sublevels $| \pm 1 \rangle $. An applied magnetic field $ B_0 $ along the NV axis splits the states $| +1 \rangle $ and $| -1 \rangle $. We consider the states $| 0 \rangle $ and  $| +1 \rangle $ only when manipulating the state of the $^{13}$C nuclear spin.
 With the nuclear Zeeman interaction and the hyperfine interactions of two spins, the energy levels of the system split to four. And the transition frequencies of these states are labelled with $f_{1}$, $f_{2}$, $f_{rf}$, as shown in the (Fig. \ref{figure1}).

The $^{13}$C nuclear spin is highly polarized into $ | \uparrow \rangle $ state under the applied magnetic field of 515 Gauss and the laser pumping, as shown in (Fig. \ref{figurecw} A). With a radio frequency $ \pi $ pulse with frequency $f_{rf}$ to transfer population between the $^{13}$C nuclear spin state $ | \uparrow \rangle $ and $ | \downarrow \rangle $ in  NV's $| 0 \rangle $ subspace, the nuclear spin is prepared to $ | \downarrow \rangle $ state (Fig. \ref{figurecw}B). Polarization of $^{13}$C nuclear spin  can be manipulated by controlling the rotation angle ($\theta$) of RF pulse (Fig. \ref{fig:rf}).

To measure the population of $^{13}C$ nuclear spin at $ | \uparrow \rangle $ and $ | \downarrow \rangle $ states, we adopted the Ramsey sequence as shown in the main text.
The length of the microwave $ \pi $ pulse measured at sensor spin’s resonance frequency between $\left | m_s=0, m_I=\uparrow \right >$ and $\left | m_s=-1, m_I=\uparrow \right >$
(f1) is 234 ns.
The length of the microwave $ \pi $ pulse measured at resonance frequency between $\left | m_s=0, m_I=\downarrow \right >$ and $\left | m_s=-1, m_I=\downarrow \right >$ (f2) is 154 ns.
The Rabi oscillations at this two frequencies are shown in the Fig.\ref{fig:rabi}A and B, respectively.
Limited by the power of the microwave, the whole magnetic field information is unavailable in a single Ramsey sequence.
Therefore, we adopt the strategy to split the sensing process into two steps to record the sensor spin's evolution.
In the first step, MW1 is detuned from f1 by +1MHz (4321.0 MHz).
The large off-resonance between the frequency of MW1 the resonance frequency f2 leaves only the information of $\left | m_I=\uparrow \right >$ state observable.
Complementarily, the procedure above is repeated but with MW2 detuned from f2 by -1MHz (4305.5 MHz ).
A typical experimental result is shown in (Fig. \ref{fig:time}) with the RF rotation angle $ \theta = 9/14 \pi$.

\subsection{Data processing methods}

To obtain the magnetic field distribution $f(B)$, we transform the signal from time domain to frequency domain and divide it by electron's gyromagnetic ratio.
In the experiment,
the effective magnetic field $\delta B$ and the fluctuation $\delta B^2$ are obtained
 by adding up the value of data point over the given interval of [-$B_{th}$,$B_{th}$], where $B_{th}$ is chosen to be 0.178 Gauss (the whole interval is approximately 2$\sigma$) in this experiment.

\begin{eqnarray}
\delta B = \frac{1}{\mathcal{N}} \sum_{-B_{th}}^{B_{th}}B\cdot f(B),
\end{eqnarray}
\begin{eqnarray}
\delta B^2 = \frac{1}{\mathcal{N}}\sum_{-B_{th}}^{B_{th}}B^2\cdot f(B) -{(\delta B)}^2,
\end{eqnarray}
where $\mathcal{N}$ is normalization factor for the distribution $f(B)$.
The theoretical values of the mean magnetic field and magnetic field fluctuation are calculated by the density matrix of the target spin.
\begin{eqnarray}
\delta B = \rho_{11} \cdot B_\uparrow+\rho_{22} \cdot B_\downarrow
\end{eqnarray}
\begin{eqnarray}
\delta B^2 = (\rho_{11} \cdot B_\uparrow)^2+(\rho_{22} \cdot B_\downarrow)^2-(\delta_B)^2
\end{eqnarray}
where $B_\uparrow~(B_\downarrow)$ is the effective magnetic field applied by the $^{13}C$ spin up (down) state.

The natural polarization at 515 Gauss is measured as
\begin{equation}
P_0=B_{exp}/ B_{\uparrow},
\end{equation}
where $B_{exp}$ is the measured effective magnetic field without RF pulse.

The error bar is obtained by dividing the data into three groups. All error bars are supposed to be such that $x\pm \delta x$ is a 68.3$\%$ confidence interval.

\subsection{Simulation of the experiment}

All the parameters of the simulation are set to be same as those in the experiments.
The experimental decoherence time $\mathrm{T}_{2}^{*}$ of the sensor spin is 1.8 $\mu s$.
An example of simulation result is shown in (Fig. \ref{fig:time}) which is in agreement with the experimental results.

It is worth noting that when the nuclear spin is fully polarized to spin up (down) state, i.e., the polarization is set to be $\pm 1$ in the simulation, there is still a small nonzero peak at the other state. It is caused by the power broadening of detecting MW and minor aliasing effect (Fig. \ref{fig:theory0}). This nonzero peak has little impact on our experiments but in the fully polarized scenario because this peak is higher than the white noise, leaving it the main contribution of the magnetic fluctuation. Since the origin of the peak is trivial, we can get rid of this trivial fluctuation by subtracting its contribution when processing the experiment data according to the simulation result.

\newpage

\newpage
\begin{table}[htbp]
\label{table1}
  \resizebox{\textwidth}{!}{
\centering
\begin{tabular}{|p{4cm} | p{2.2cm} | p{2.8cm} | p{3cm} | p{2cm} |p{2cm}|}
\hline
\textbf{Typical spectroscopy
Methods}& \textbf{Mechanism}& \textbf{Information}& \textbf{Revealing Symmetry} & \textbf{damage to material}& \textbf{Large Device Requirement}\\
\hline
photon scattering &	Real particle interaction&	Crystalline structure and electronic band structure	&indirect&	No&\\	
\hline
neutron scattering &Real particle interaction&Crystalline Structure&indirect&Yes&Nuclear reactor\\
\hline
Electron diffraction&Real particle interaction&Crystalline structure&indirect&Yes&\\	
\hline
muon-spin rotation spectroscopy&Real particle interaction&Time-reversal symmetry&direct but only for Time-reversal symmetry&Yes&Large Accelerator\\
\hline
Quantum Atmosphere &Virtual-particle exchange process&P, T symmetry and other hidden symmetry&direct&No&\\	
\hline
\end{tabular}
}
\caption{Comparison between traditional spectroscopy and QA-based spectroscopy}
\label{table1}
\end{table}

\newpage
\begin{figure*}[htbp]\centering
\includegraphics[width=0.9\columnwidth]{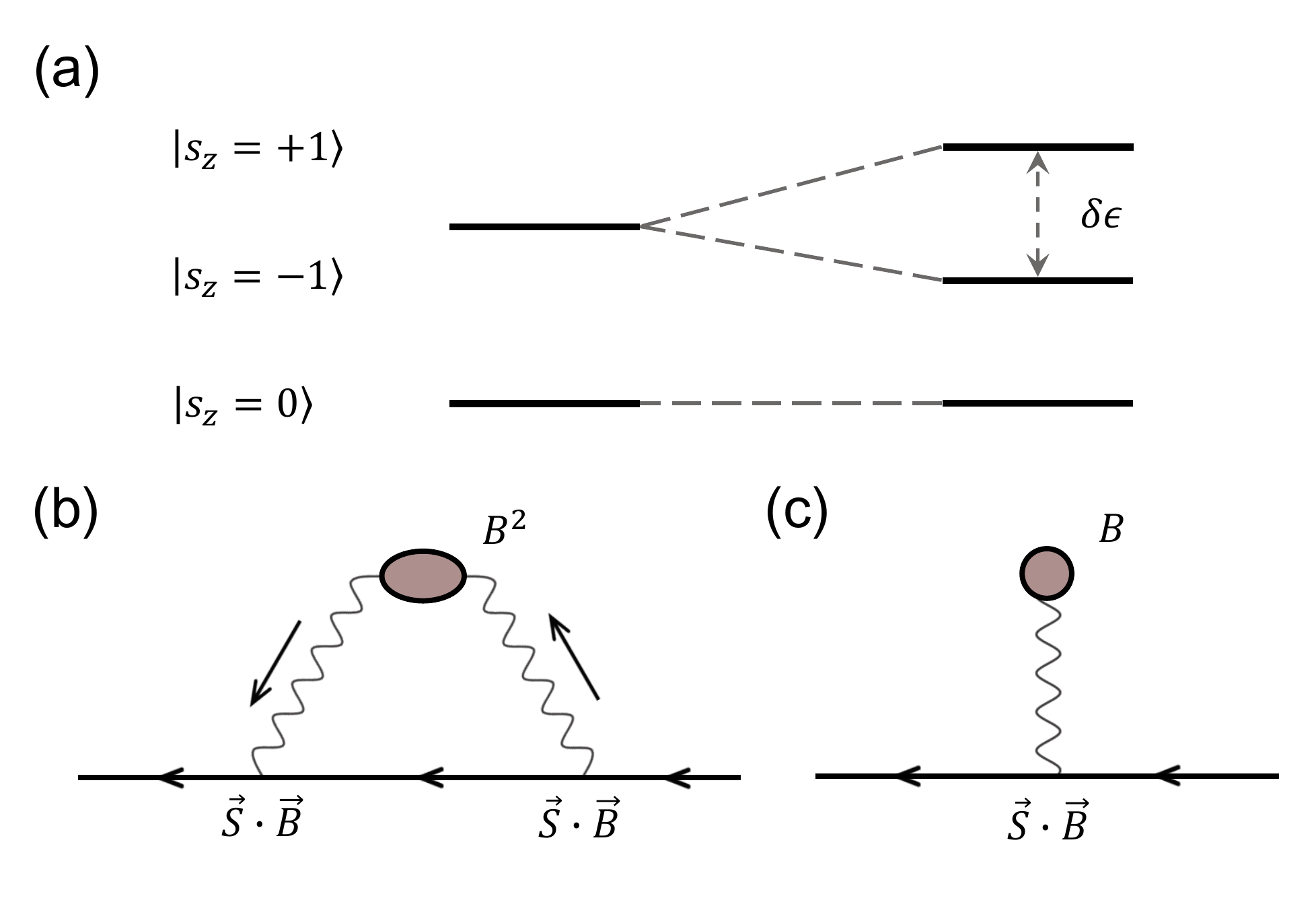}
\caption{\label{figure_Feynman} \textbf{Time-reversal symmetry in terms of general theory of quantum atmosphere.} (a) The upper panel shows that the spectra of the sensor is modified in the presence of the material. The energy level of the spin-up state and spin-down state will split with an energy difference $\delta \epsilon$ when a spin is put in a time-reversal symmetry broken atmosphere
The lower panel shows different symmetry in terms of Feynman diagram.
The black straight (wavy) line with an arrow represents propagator for sensor spin (photons). The grey circle represents the scattering vertex $\Pi(p\prime-p)$ where the symmetry information in a material is encoded. (b) shows the two-photon-exchange process with conserved time-reversal symmetry. (c) shows the first order process when the time-reversal symmetry is broken.
}
\end{figure*}

\newpage
\begin{figure*}[htbp]\centering
\includegraphics[width=0.9\columnwidth]{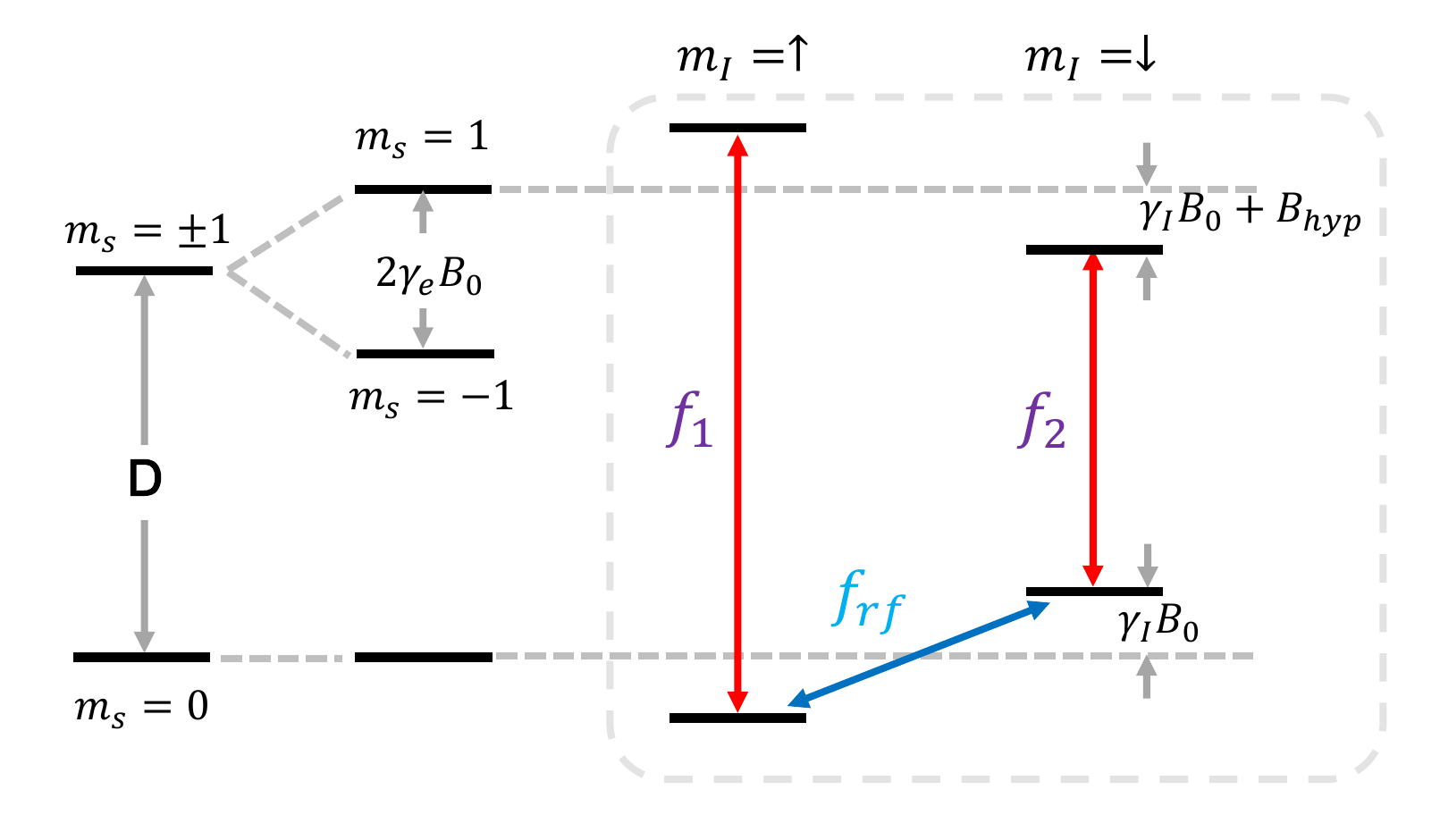}
\caption{\label{figure1} \textbf{Energy levels of the system.} Zero field splitting $D=2.87$ GHz separates the ground state $\left | S_z = 0 \right >$ and $\left | S_z = \pm1 \right >$ of the NV center's electron spin.  The $\left | S_z = \pm1 \right >$ states experience a Zeeman shift in the presence of the applied magnetic field $B_{0}$, $\gamma_e=2.8 $MHz/Gauss is the electron spin gyromagnetic ratio. The hyperfine interaction with $^{13}C$ nuclear spin splits each state into two, corresponding to different $^{13}C$ nuclear spin state  $\left | m_I = \uparrow \right >$ and $\left | m_I = \downarrow \right >$. $B_{hyp}$ denotes the coupling between electron and nuclear spins. $\gamma_I=1.07 $kHz/Gauss is $^{13}C$ nuclear spin gyromagnetic ratio. $f_{1}$ and $f_{2}$ are the transition frequencies of electron spin between $\left | S_z = 0 \right >$ and $\left | S_z = +1 \right >$, corresponding to $^{13}C$ in the $\left | m_I = \uparrow \right >$ and $\left | m_I = \downarrow \right >$ respectively. $f_{rf}$ is the transition frequency of $^{13}C$ nuclear spin in the subspace of electron spin's  $\left | S_z = 0 \right >$ state.
 }
\end{figure*}

\newpage
\begin{figure*}[htbp]\centering
\includegraphics[width=0.7\columnwidth]{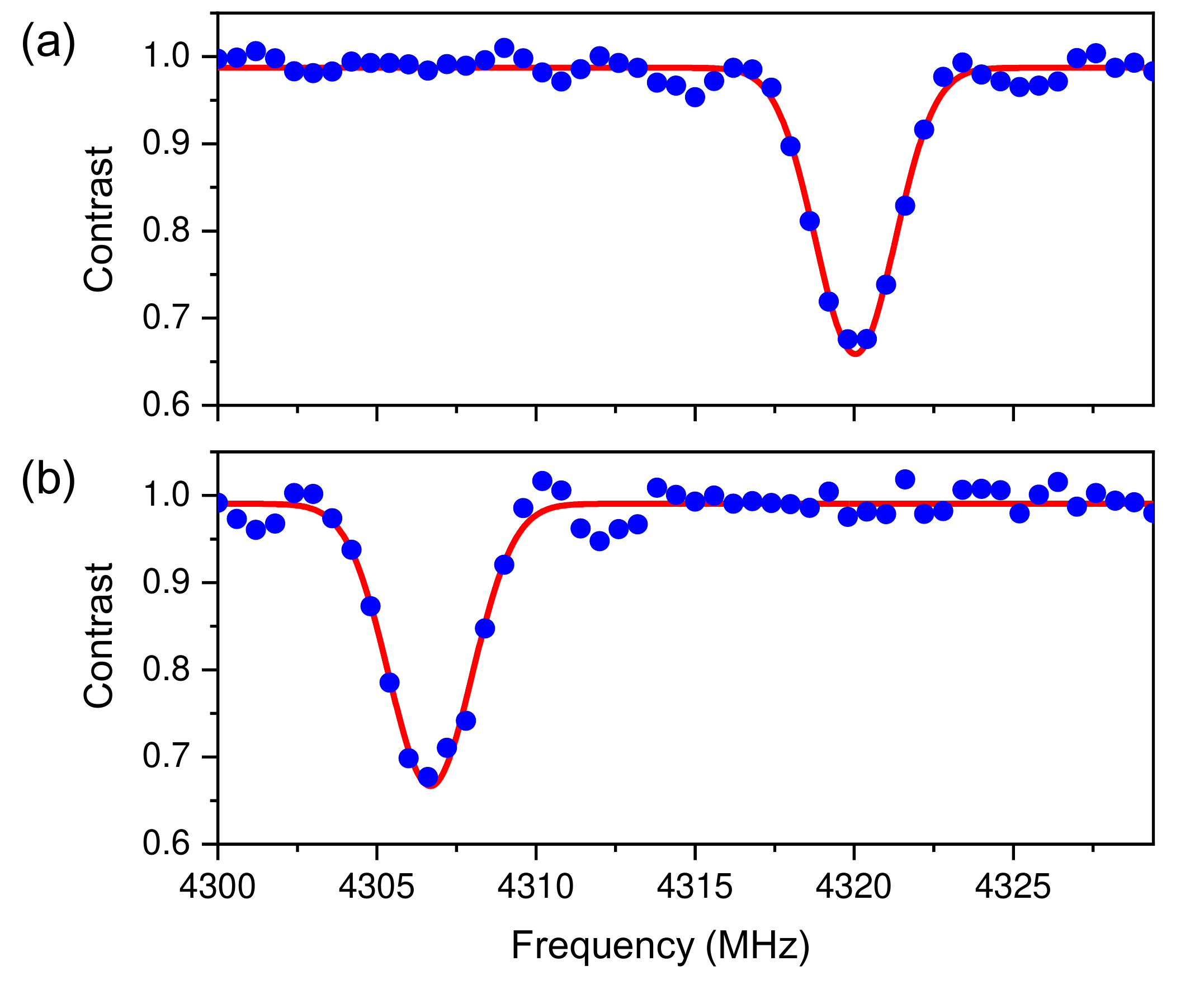}
\caption{\label{figurecw} \textbf{ODMR spectra of the NV center in different $^{13}C$ nuclear spin state.} (a)The ODMR spectrum at 515 Gauss without RF manipulation. There is only a peak at 4320.0 MHz corresponding to $^{13}C$'s $ | \uparrow \rangle $ state. (b) After a RF $ \pi $ pulse of frequency $f_{rf}$, the nuclear spin is pumped to $ | \downarrow \rangle $ state, the ODMR spectrum only shows a peak at 4306.5 MHz. Coupling between NV and the $^{13}C$ nuclear spin can be obtained  from these two spectrum.
 }
\end{figure*}

\newpage
\begin{figure*}[htbp]\centering
\includegraphics[width=0.9\columnwidth]{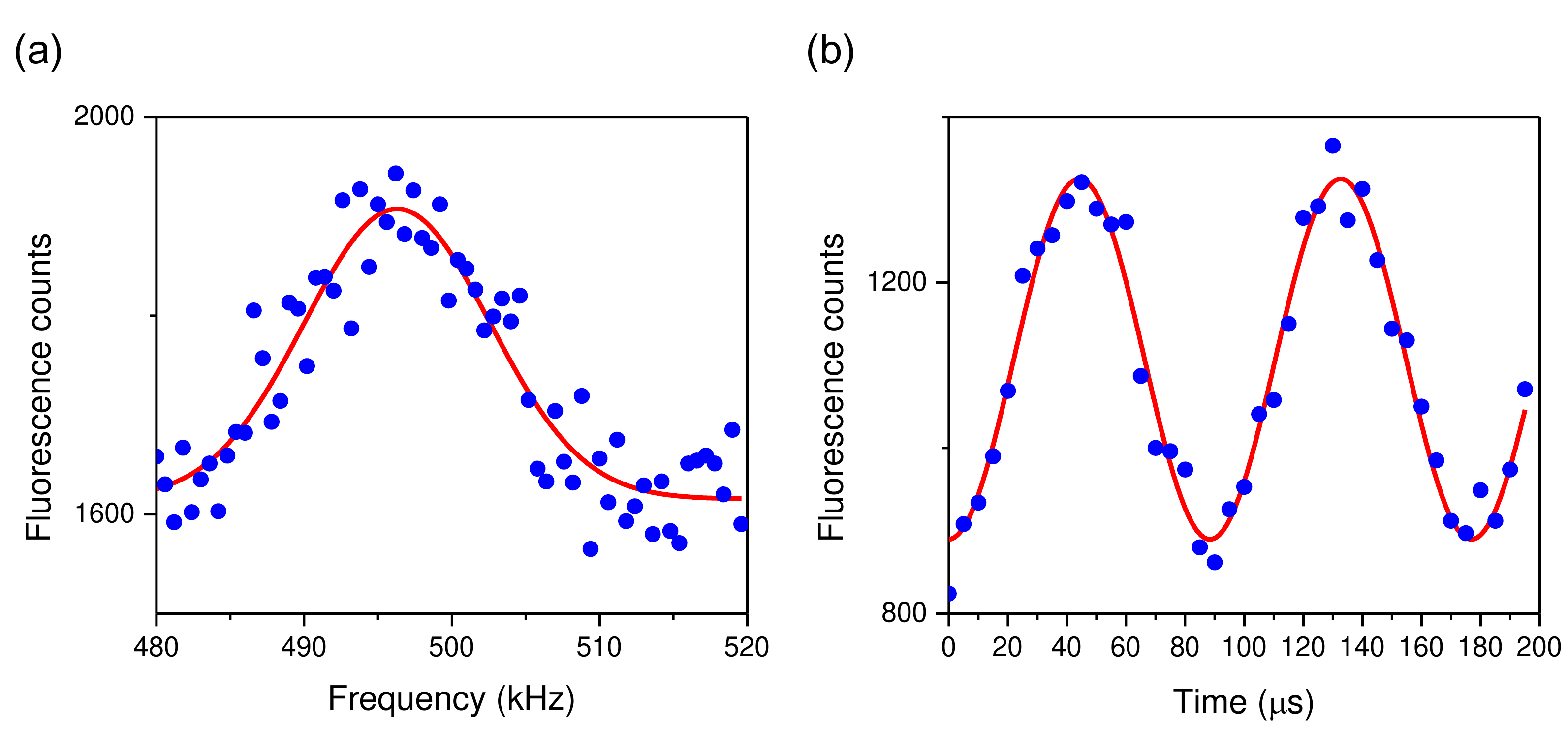}
\caption{\textbf{The Rabi oscillations of the nuclear spin.} (a) The spectrum of $^{13}C$ nuclear spin. The transition frequency of $^{13}C$ nuclear spin is measured to be 496 kHz by sweeping the frequency of RF pulses. (b) The nutation of the $^{13}C$ nuclear spin. The length RF $ \pi $ pulse is measured to be 45 $\mu s$ by varying the duration of the resonant RF pulses. These two steps are alternatively iterated to obtain more precise resonant frequency and $ \pi $ duration.
 }\label{fig:rf}
\end{figure*}

\newpage
\begin{figure*}[htbp]\centering
\includegraphics[width=0.9\columnwidth]{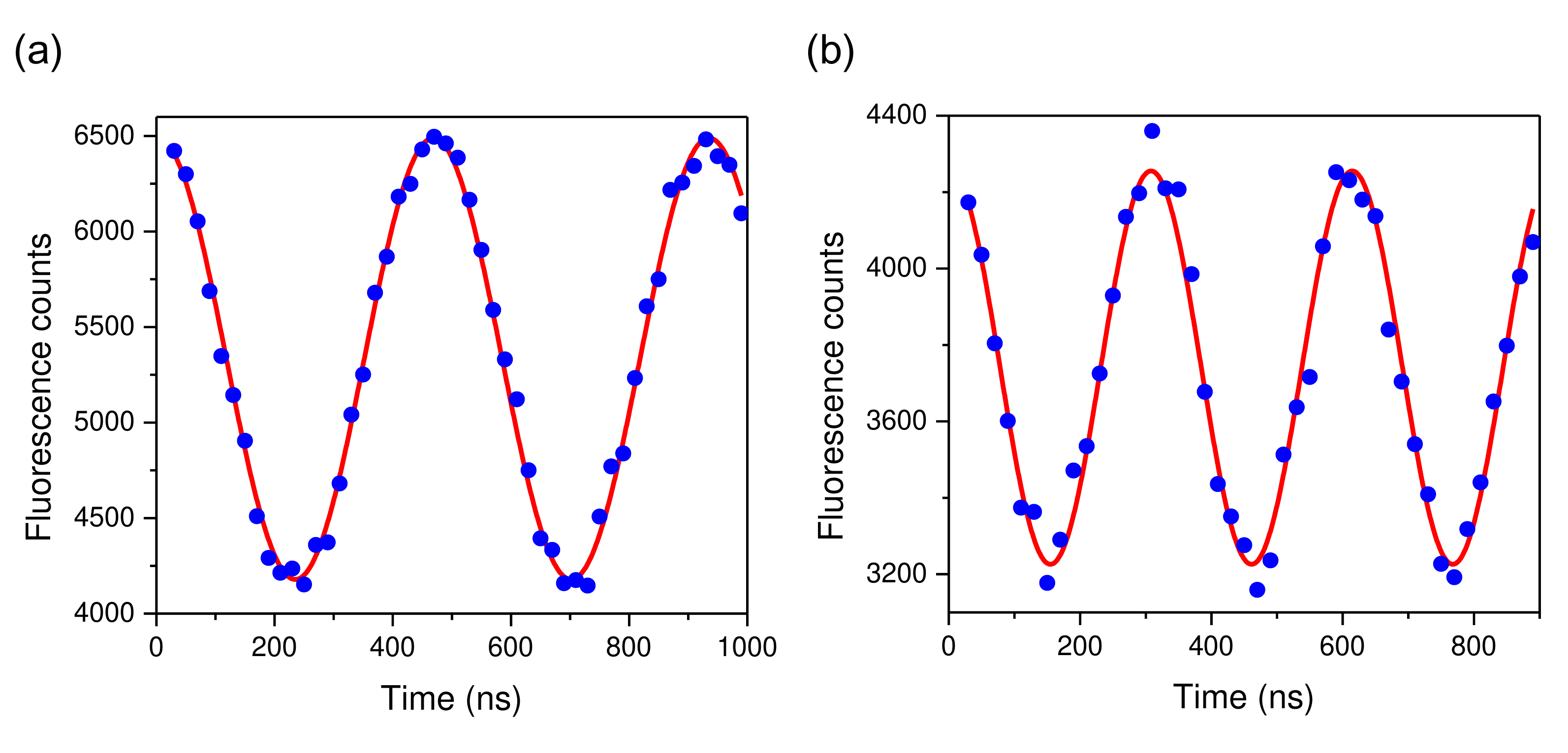}
\caption{\textbf{The Rabi oscillations of the sensor spin.} (a) MW1 is set to be detuned from $f_{1}$ by +1MHz (a typical value 4321.0 MHz) to detect the population of $^{13}C$ at $ | \uparrow \rangle $ state.  (b) MW2 is detuned form $f_{2}$ by -1MHz (a typical value 4305.5 MHz) to detect the population of $^{13}C$ at $ | \downarrow \rangle $ state. The fluorescence counts of each diagram shows the relative population of $^{13}C$ nuclear spin's $ | \uparrow \rangle $ and $ | \downarrow \rangle $ state.
 }\label{fig:rabi}
\end{figure*}

\newpage
\begin{figure*}[htbp]\centering
\includegraphics[width=0.9\columnwidth]{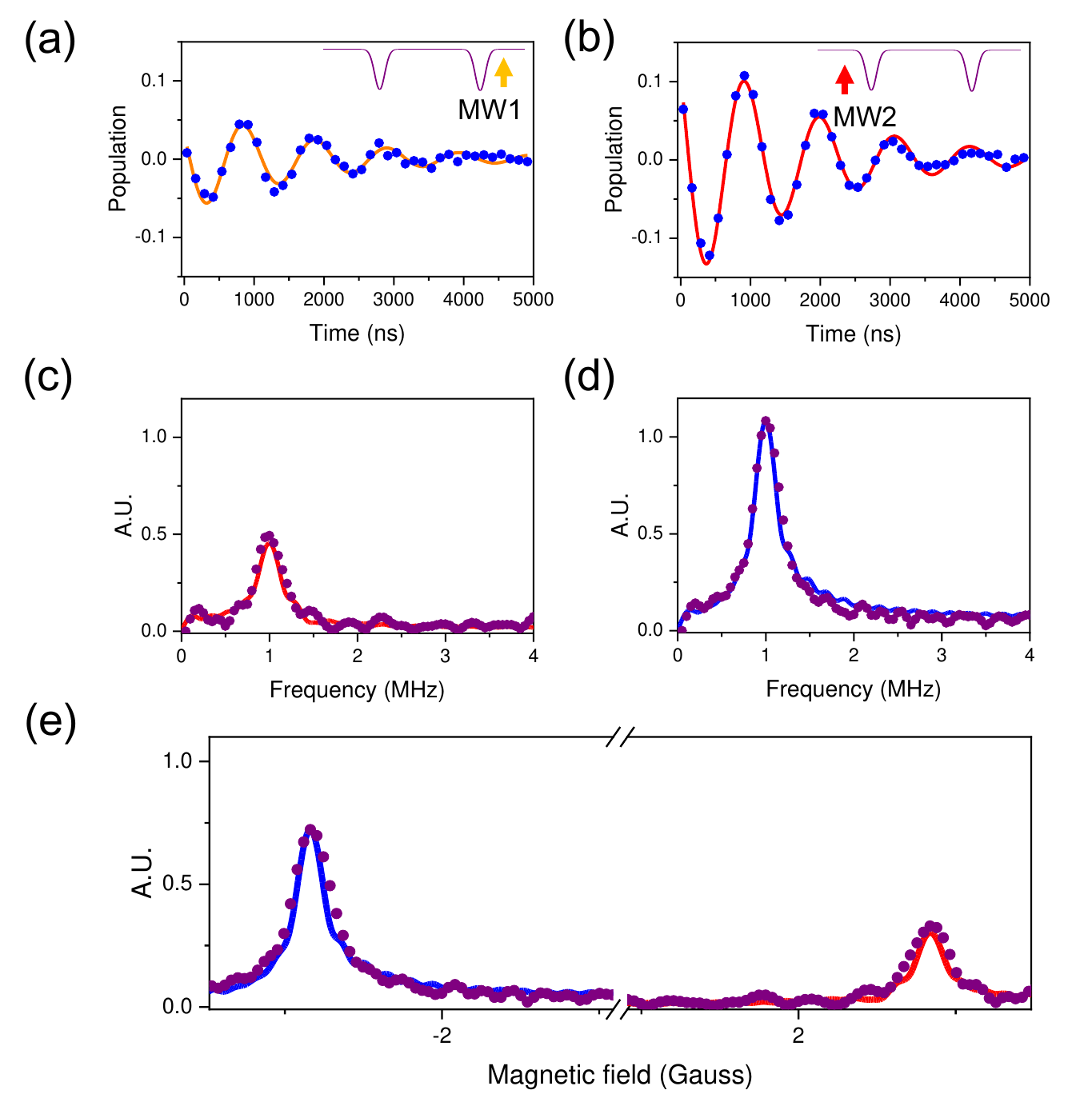}
\caption{\textbf{The two-step Ramsey sequence.} (a)(b) The time domain results with RF rotation angle $ \theta = 9/14 \pi$. The left (right) panel is the time domain signal obtained by sensing sequence with MW1 (MW2). (c)(d) The frequency domain result by Fourier transformation of the data in (a) and (b). The scattered dots are the experimental results while the real lines are our simulant results.(e) The typical time-reversal symmetry broken pattern has been successfully observed in the  $^{13}C$ nuclear spin's quantum atmosphere.
 }\label{fig:time}
\end{figure*}

\newpage
\begin{figure*}[htbp]\centering
\includegraphics[width=0.9\columnwidth]{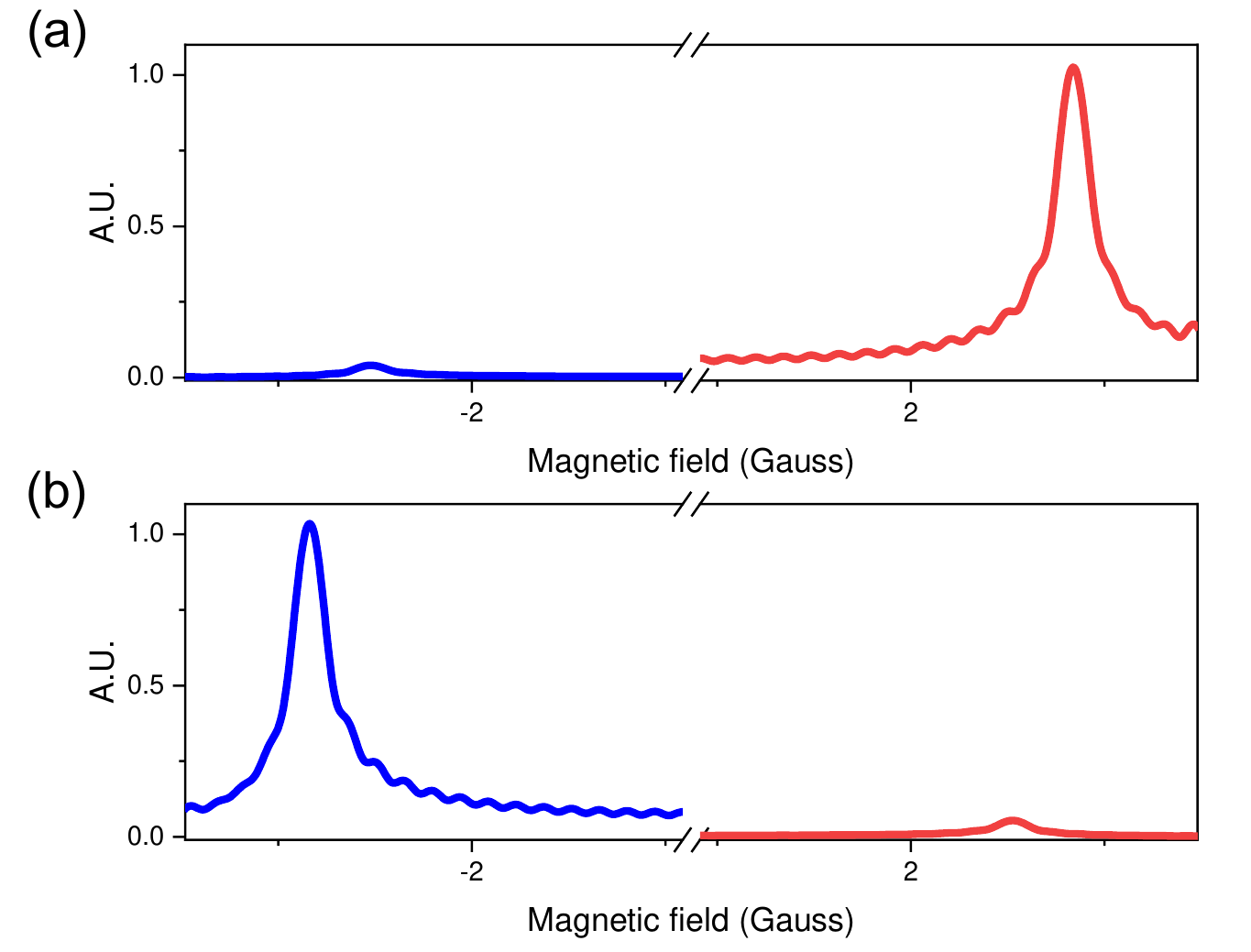}
\caption{\textbf{Simulations in completely polarized scenario.} (a) The simulant spectra when the $^{13}C$ nuclear spin is fully polarized into $ | \uparrow \rangle $ state. (b) The simulant spectra when the $^{13}C$ nuclear spin is fully polarized into $ | \downarrow \rangle $ state. There is a small peak in the unpolarized state resulting from the power broadening of detecting MW.
}\label{fig:theory0}
\end{figure*}

\end{document}